\documentclass[pre,aps,twocolumn,superscriptaddress,showpacs]{revtex4-1}
\usepackage{amssymb}
\usepackage{epsfig}
\usepackage{amsmath}
\usepackage{subfigure}
\usepackage{graphicx}
\usepackage{textcomp}
\usepackage{url}
\usepackage{float}
\usepackage{color}
\usepackage{cases}
\usepackage[normalem]{ulem}

\usepackage{enumitem} 

\usepackage[colorlinks=true, urlcolor=blue, anchorcolor=blue, citecolor=blue,filecolor=blue,linkcolor=blue,menucolor=blue
]{hyperref}

\usepackage{color}
\definecolor{Blue}{rgb}{0.00, 0.00, 0.80}
\definecolor{Red}{rgb}{0.80, 0.00, 0.00}
\definecolor{Green}{rgb}{0.00, 0.50, 0.00}

\usepackage{fixmath}
\newcommand{\vect}[1]{\mathbold {#1}} 

\newcommand{\x}{\vect{x}}

\usepackage{epsfig}
\usepackage{textcomp}
\usepackage{epstopdf}

\newcommand{\nn}{\nonumber}
\newcommand{\be}{\begin{equation}}
\newcommand{\ee}{\end{equation}}
\newcommand{\bea}{\begin{eqnarray}}
\newcommand{\eea}{\end{eqnarray}}

\begin{document}

\title{Full distribution of the ground-state energy of potentials with weak disorder}

\author{Naftali R. Smith}
\email{naftalismith@gmail.com}
\affiliation{Racah Institute of Physics, Hebrew University of Jerusalem, Jerusalem 91904, Israel}
\affiliation{Department of Environmental Physics, Blaustein Institutes for Desert Research, Ben-Gurion University of the Negev, Sede Boqer Campus, 8499000, Israel}


\begin{abstract}

We study the full distribution $P(E)$ of the ground-state energy of a single quantum particle in a potential $V(\x) = V_0(\x) + \sqrt{\epsilon} \, v_1(\x)$, where $V_0(\x)$ is a deterministic ``background'' trapping potential and $v_1(\x)$ is the disorder. 
We consider arbitrary trapping potentials $V_0(\x)$ and white-noise disorder $v_1(\x)$,  in arbitrary spatial dimension $d$.
In the weak-disorder limit $\epsilon \to 0$, we find that $P(E)$ scales as $P(E) \sim e^{-s(E)/\epsilon}$. The large-deviation function $s(E)$ is obtained by calculating the most likely configuration of $V(\x)$ conditioned on a given ground-state energy $E$. 
For infinite systems, we obtain $s(E)$ analytically in the limits $E \to \pm \infty$ and $E \simeq E_0$ where $E_0$ is the ground-state energy in the absence of disorder. We perform explicit calculations for the case of a harmonic trap $V_0(\x) \propto x^2$  in dimensions $d\in\left\{ 1,2,3\right\}$.
Next, we calculate $s(E)$ exactly for a finite, periodic one-dimensional system with a homogeneous background $V_0(x)=0$. We find that, remarkably, the system exhibits a sudden change of behavior as $E$ crosses a critical value $E_c < 0$: At $E>E_c$, the most likely configuration of $V(x)$ is homogeneous, whereas at $E < E_c$ it is inhomogeneous, thus spontaneously breaking the translational symmetry of the problem.
As a result, $s(E)$ is nonanalytic: Its second derivative jumps at $E=E_c$. We interpret this singularity as a second-order dynamical phase transition.

\end{abstract}

\maketitle

\section{Introduction}

The analysis of a single quantum particle affected by a disordered potential has attracted interest for decades \cite{Halperin65, HalperinLax66, ZL66, VV95, BW12, RoatiEtAl08, ArnoldEtAl16, CWZ22, GH24}, with applications to Anderson localization \cite{RoatiEtAl08, ArnoldEtAl16}, semiconductors \cite{VanMieghem92, GH24}, the quantum Hall effect \cite{Affleck83}, photovaoltaic absorption \cite{JeanEtAl17,WOA21} and many more. Of central interest is the effect of the disorder on the energy spectrum, but other properties such as tunnelling amplitudes \cite{FPY96, DaCostaEtAlExperiment2000, Luck04, KRL08} and dynamics \cite{BillyEtAl08} have attracted lots of attention too.

Much of the work has focused on the density of states (DOS). The exact DOS has been known for one-dimensional, translationally invariant systems with Gaussian disorder for some time \cite{Halperin65, VV95}. The DOS in the low-lying regime is dominated by rare configurations of the disorder, and thus can be calculated using the optimal fluctuation method (OFM), or instanton method.
Such calculations have been carried out in a variety of settings \cite{HalperinLax66, ZL66, Affleck83, Yaida16, GH24}, including dimensions $d=1$ \cite{ZL66}, $d=1,2,3$ with subleading corrections \cite{GH24} and in $d=2$ with a magnetic field
\cite{Affleck83}.

How is the energy spectrum of a non-translationary invariant system affected by disorder? In this paper, we make a first step towards answering this question, by studying the distribution $P(E)$ of the ground-state energy $E$ in a potential of the structure
$V(\x) = V_0(\x) + \sqrt{\epsilon} \, v_1(\x)$,
where $V_0(\x)$ is a given trapping potential and $v_1(\x)$ represents the disorder.
The effect of disorder on the ground-state energy has been studied for the particular case of ``Bernoulli'' disorder (in which the potential randomly alternates in space between two values) in a finite system corresponding to a ``particle-in-a-box'' potential $V_0(\x)$ \cite{BW12, CWZ22}. 

In the present work, we consider general trapping potentials $V_0(\x)$. 
In order to make analytic progress, we focus on the physically-relevant case in which the disorder is modeled as white noise. Moreover, we consider the limit of weak disorder intensity $\epsilon \to 0$. In this limit, the problem naturally falls within the realm of large-deviation theory \cite{Varadhan,O1989,DZ,Hollander, T2009,Derrida11, bertini2015, MS2017,Touchette2018, Derrida2007,   Baek15, Baek17, Baek18,  Shpielberg2016, SKM2018, ZM16, SMS2018,KLP18, SGG21,NT22, Agranov2020, Smith22KPZ, Smith22Chaos, ACJ22, SGG23, ACJ23, ME23, SSM23}, 
since any value $E \ne E_0$ becomes a rare event, where $E_0$ is the ground-state energy in the absence of disorder, i.e., the ground-state energy for the potential $V(\x)=V_0(\x)$.
Furthermore, at $\epsilon \to 0$, $P(E)$ is dominated by the most likely realization of $V(\x)$, conditioned on the value of $E$. 
This enables us to use the OFM \cite{HalperinLax66,ZL66,Lifshitz}, a method which is known to be general and versatile, and has been applied to many physical systems, including turbulence \cite{turb1}, stochastic lattice gases \cite{bertini2015} stochastic reactions \cite{EK,MS2011}, non-equilibrium surface growth \cite{MKV}. The OFM has also been applied successfully in the study of disordered systems such as particles diffusing in random media \cite{LV01, Meerson22,VLM24}, besides the applications mentioned above for the $E \to -\infty$ tail of the DOS in quantum systems.


The generality of the OFM enables us to consider a relatively broad range of scenarios [in particular, general trapping potentials $V_0(\x)$]. Its application to the calculation of $P(E)$ immediately yields the scaling behavior
$P(E) \sim e^{-s(E)/\epsilon}$. The large-deviation function $s(E)$ is calculated by solving an optimization problem which involves finding the most likely realization of the potential $V(\x)$ conditioned on $E$. The optimization problem takes the form of an underlying classical Hamiltonian system.

The remainder of the paper is arranged as follows.
In Section \ref{sec:Model} we define the problem of the statistics of the ground-state energy for a single particle in a disordered potential, and formulate the OFM for this problem.
In Section \ref{sec:OFMsol}, we solve the OFM problem analytically  for the general case in the limits $E \! \to \! \pm\infty$ and $E \simeq E_0$. 
To illustrate our results, we give explicit numerical and analytic results for the case in which the deterministic part of the potential is harmonic, $V_0(\x) \propto x^2$ (where here and below $x = \left|\x \right|$),  in dimensions $d\in\left\{ 1,2,3\right\}$.
In Section \ref{sec:PBC}, we solve the OFM problem exactly for a finite, one-dimensional, periodic, homogeneous [$V_0(x)=0$] system. We uncover a singularity of $s(E)$ for this case, which we interpret as a second-order dynamical phase transition (DPT) \cite{footnote:DPT}.
In Section \ref{sec:discussion} we briefly summarize and discuss our main findings.

%
%

%




%



\section{Model and OFM formalism}
\label{sec:Model}

We consider a single quantum particle in $d$ dimensions, in a potential of the form 
\be
\label{Vdef}
V(\x) = V_0(\x) + \sqrt{\epsilon} \, v_1(\x),
\ee
where $V_0(\x)$ is a given, deterministic trapping potential and  $V_1(x) = \sqrt{\epsilon} \, v_1(\x)$ represents the disorder.
Disordered potentials \eqref{Vdef} have been extensively studied in the context of quantum systems,  mostly for the particular case of translationally-invariant systems, $V_0(\x)=0$ \cite{VV95, Yaida16, GH24}, or for finite systems \cite{Halperin65,BW12, CWZ22} e.g., in which $V_0(\x)$ describes a particle in a box.
Disordered potentials have also been studied in the context of diffusion in random media \cite{Zwanzig88, LV01, Meerson22,VLM24}.
We assume that the disorder takes the form of a white Gaussian noise with 
$\left\langle v_{1}\left(\x\right)\right\rangle =0$
and
$\left\langle v_{1}\left(\x\right)v_{1}\left(\x'\right)\right\rangle =\delta\left(\x-\x'\right)$ \cite{Halperin65, VV95},
where here $\delta(\cdots)$ denotes the $d$-dimensional Dirac delta function.
We are interested in the distribution $P(E)$ of the ground-state energy $E$.

In order to make analytic progress, we focus here on the limit in which the disorder intensity is weak, $\epsilon \to 0$. In this limit, typical realizations of the disorder will lead to ground-state energies $E  \simeq E_0$, where $E_0$ is the ground-state energy of the potential $V_0(\x)$ alone. However, one can ask how the disorder affects the ground-state energy, and we study this question here by investigating the full distribution $P(E)$, both for typical fluctuations $E\simeq E_0$ as well as large deviations, i.e., when $E$ is not close to $E_0$.

The starting point of the OFM formulation is the probability (density) of a given realization of the (white-noise) disorder,
\be
\mathcal{P}\left[v_{1}\left(\x\right)\right]\sim e^{-\frac{1}{2}\int v_1\left(\x\right)^{2}d\x} \, ,
\ee
where here and below the integration is over the entire $d$-dimensional space.
Using this with Eq.~\eqref{Vdef} one finds that, up to a Jacobian that is subleading in the limit $\epsilon \to 0$, the probability for observing a given realization of $V(\x)$ is
\be
\mathcal{P}\left[V\left(\x\right)\right]\sim e^{-\frac{1}{2\epsilon}\int \left[V\left(\x\right)-V_{0}\left(\x\right)\right]^{2}d\x} \, .
\ee
One can formally write $P(E)$ in the form of a path integral, by summing the contributions that originate in realizations of the disorder $V_1(\x)$ for which the ground-state energy of the potential $V(\x)$ equals $E$. In the limit $\epsilon \to 0$, this path integral can (in the leading order) be evaluated by using the saddle-point approximation. This immediately yields the scaling behavior
\be
\label{LDP}
P\left(E\right)\sim e^{-s\left(E\right)/\epsilon} \, ,
\ee
where the large-deviation function $s(E)$ is given by evaluating the minimum ``action''
\be
\label{sdef}
s=s\left[V_{1}\left(\x\right)\right]=\frac{1}{2}\int V_{1}\left(\x\right)^{2}d\x
\ee
on the ``optimal'' (or most likely) realization of $V_1(\x)$ constrained on the ground-state energy of the potential $V(\x) = V_0(\x) + V_1(\x)$ being equal to $E$.
The latter constraint is incorporated by minimizing the modified action functional
\be
s_{\lambda}\left[V_{1}\left(\x\right)\right]=s\left[V_{1}\left(\x\right)\right]-\lambda E\left[V_{1}\left(\x\right)\right]
\ee
where $E\left[V_{1}\left(\x\right)\right]$ is the ground-state energy as a functional of the disorder, and $\lambda$ is a Lagrange multiplier.

The action functional \eqref{sdef} has a relatively simple form. However, the constraint on a given $E$ makes matters complicated as it is difficult to write $E = E\left[V_{1}\left(\x\right)\right]$ explicitly as a functional of $V_1(\x)$. This difficulty is cirumvented if one formulates the problem in terms of the ground-state wave function  $\bar{\psi}(\x)$ that corresponds to the optimal configuration of $V_1(\x)$,  and is normalized $\int \bar{\psi} (\x)^2 d\x = 1$. Let $\delta V_1(\x)$ be a small variation in the disorder. Then the corresponding variation in the  modified action is
\be
\delta s_{\lambda}=\delta s-\lambda\delta E \, .
\ee
Here
\be
\delta s=\int V_{1}\left(\x\right)\delta V_{1}\left(\x\right)d\x+O\left(\delta V_{1}^{2}\right) 
\ee
and, from first-order perturbation theory of quantum mechanics, one finds that the variation of the ground-state energy is given by 
\be
\delta E=  \int\bar{\psi}\left(\x\right)^{2}\delta V_{1}\left(\x\right)d\x+O\left(\delta V_{1}^{2}\right)\,.
\ee

For the optimal realization of the disorder, the variational derivative $\delta s_{\lambda}/\delta V_1$ must vanish (for arbitrary variations $\delta V_1$). This leads to the relation
\be
\label{V1lambdapsibar}
V_{1}\left(\x\right)= \lambda\bar{\psi}\left(\x\right)^{2} \, .
\ee
It is convenient to work with the \emph{unnormalized} wavefunction
\be
\psi\left(\x\right)=\sqrt{\left|\lambda\right|} \, \bar{\psi}\left(\x\right) \, ,
\ee
so that
\be
\label{V1ofPsi}
V_1(\x) = \pm \psi(\x)^2 \, ,
\ee
where the choice of sign is as follows:
For $E>E_0$, the disorder increases the ground-state energy and therefore $V_1 > 0$, and vice versa for $E<E_0$. Thus the sign in Eq.~\eqref{V1ofPsi} equals  $\text{sgn}(E-E_0)=\text{sgn}(\lambda)$ (in particular, note that $\lambda$ vanishes at $E=E_0$).
On the other hand, $\psi(\x)$ satisfies the time-independent Schr\"{o}dinger equation (choosing units such that $\hbar^2 / m = 1$)
\be
-\frac{1}{2} \nabla^2 \psi\left(\x\right)+V\left(\x\right)\psi\left(\x\right)=E\psi\left(\x\right) \, .
\ee
Putting the last two equations together, we obtain the equation
\be
\label{psiEq}
-\frac{1}{2}\nabla^2 \psi\left(\x\right)+V_{0}\left(\x\right)\psi\left(\x\right) + \text{sgn}(E-E_0)\psi\left(\x\right)^{3}=E\psi\left(\x\right)
\ee
which is to be solved subject to the boundary conditions $\psi \to 0$ as $x \to \infty$ that follow from normalizability.

Eq.~\eqref{psiEq} is a nonlinearly-modified Schr\"{o}dinger-like equation. 
After solving Eq.~\eqref{psiEq} for given $V_0(\x)$ and $E$, one obtains the large-deviation function $s(E)$ by plugging Eq.~\eqref{V1ofPsi} into \eqref{sdef}:
\be
\label{sOfPsi}
s=\frac{1}{2}\int \psi\left(\x\right)^{4}d\x \, .
\ee
Eq.~\eqref{V1ofPsi} also yields the optimal realization of the disorder $V_1(\x)$ conditioned on $E$.

\section{Solving the OFM problem in limiting cases for an infinite system}
\label{sec:OFMsol}

In $d=1$ or in $d>1$ assuming rotational symmetry, Eq.~\eqref{psiEq} can be solved numerically by using the shooting method, which reduces the boundary value problem to an initial value problem,  see e.g. Ref.~\cite{SB80}. The solution $\psi(\x)$ is then plugged into Eq.~\eqref{sOfPsi}. This yields a numerical scheme for obtaining the large deviation function $s(E)$ at all values of $E$.

To make analytic progress, one must make further assumptions. In this section, we obtain $s(E)$ analytically in three limits: $E \to -\infty$, $E \to +\infty$ and $E \simeq E_0$.
For simplicity, we assume that $V_0(\x)$ has a single global minimum, which is located at the origin and that the value of $V_0$ there is $V_0(0)=0$.
Under these fairly mild assumptions, our analytic results are valid for a general trapping potential $V_0(\x)$. As an illustration, we give some explicit results for the harmonic case $V_0(\x) = \x^2 \! /2$ in dimensions $d=1,2,3$.

\subsection{$E \to -\infty$ tail}

In the tail $E \to -\infty$, the optimal realization $V_1(\x)$ of the disorder is localized around the origin and of very large magnitude, and correspondingly, the corresponding ground-state wavefunction $\psi(\x)$ is localized around the origin too. As a result, one can, in the leading order, approximate $V_0(\x)\simeq V_0(0)=0$ in Eq.~\eqref{psiEq}, yielding the simpler equation
\be
\label{psiEqNoV0}
-\frac{1}{2}\nabla^2 \psi\left(\x\right)-\psi\left(\x\right)^{3} = E\psi\left(\x\right) \, .
\ee

Eq.~\eqref{psiEqNoV0} has been encountered in the closely related context of calculating the DOS of the low-lying states \cite{GH24,Yaida16}, but also in other physical contexts such as nonlinear optics \cite{CGT64}, semiclassical theory of quantum barrier penetration \cite{Coleman77} and stochastic surface growth \cite{MSV_3d}.
This equation has been solved analytically in $d=1$ and numerically in $d=2,3$, and the action \eqref{sOfPsi} was evaluated, see e.g. Ref.~\cite{GH24}. The result (in our choice of units \cite{footnote:units}) is
\be
\label{sEMinusInf}
s\left(E\right)\simeq\begin{cases}
\frac{4\sqrt{2}}{3}\left(-E\right)^{3/2}, & d=1,\\[2mm]
-5.85E, & d=2,\\[2mm]
13.36\sqrt{-E}\,, & d=3.
\end{cases}
\ee
For completeness, we give the analytic solution for the case $d=1$ in Appendix \ref{app:EMinusInf}.

We thus find that, at $E \to -\infty$, the tail \eqref{sEMinusInf} of $P(E)$ coincides with the corresponding tail that describes the density of low-lying states $\rho(E)$ \cite{GH24}. Physically, the reason for this is that the dominant contribution to $\rho(E)$ (at $E \to -\infty$) is due to realizations of the disorder for which $E$ is the ground-state energy. Contributions to $\rho(E)$ that come from realizations for which $E$ is the energy of an excited state give subleading corrections that are beyond the accuracy of the leading-order result \eqref{sEMinusInf}.


\subsection{$E \to \infty$ tail}

For large positive $E$, the solution $\psi(\x)$ to Eq.~\eqref{psiEqNoV0} extends over the (large) spatial region $\left\{ \x\,:\,V_0\left(\x\right)<E\right\} $, and $\psi(\x)$ is negligible elsewhere. 
Within this region, the wavefunction $\psi(\x)$ changes relatively slowly in space.
As a result, the term $\nabla^{2}\psi\left(\x\right)$ in Eq.~\eqref{psiEqNoV0} is relatively small, and one can neglect it to obtain
\be
V_{0}\left(\x\right)\psi\left(\x\right)+\psi\left(\x\right)^{3}= E\psi\left(\x\right) \, ,
\ee
whose nontrivial solution is immediately found:
\be
\label{psiSolEInf}
\psi\left(\x\right)=\left(E-V_{0}\left(\x\right)\right)_{0}^{1/2}\,.
\ee
Here we denoted
\be
\left(\alpha\right)_{0}^{1/2}=\theta\left(\alpha\right)\alpha^{1/2}=\begin{cases}
0, & \alpha<0,\\[1mm]
\alpha^{1/2}, & \alpha\ge0.
\end{cases}
\ee
One can now calculate $s(E)$ (in the limit $E \to \infty$) by plugging the solution \eqref{psiSolEInf} into Eq.~\eqref{sOfPsi}, and the result is
\be
\label{sSolEInf}
s \simeq \frac{1}{2}\int\left(E-V_{0}\left(\x\right)\right)^{2}\theta\left(E-V_{0}\left(\x\right)\right)d\x\,.
\ee

Some comments are in order.
First of all, using Eq.~\eqref{V1ofPsi}, one finds that the optimal realization of the potential $V(\x)$ corresponding to the wavefunction \eqref{psiSolEInf} is
\be
V\left(\x\right)=V_{0}\left(\x\right)+V_{1}\left(\x\right)=\max\left\{ V_{0}\left(\x\right),E\right\} \, .
\ee
Thus, in the tail $E \to \infty$, the ground-state energy coincides, in the leading order, with the minimum of the potential $E \simeq \min_{\x} \! V(\x)$.
Moreover, it is interesting to notice that, for $d=1$, the wavefunction \eqref{psiSolEInf} is proportional to the Thomas-Fermi approximation for the density of a zero-temperature noninteracting fermi gas in an external potential $V_0(x)$, with fermi energy $E$, see e.g. \cite{Castin, Castin2}.

\subsection{Typical fluctuations}

For typical fluctuations, $E \simeq E_0$, it is convenient to use a slightly different approach, rather than tackling Eq.~\eqref{psiEq} directly.
 The leading order asymptotic behavior of $s(E)$ at $E \simeq E_0$ is obtained by approximating the normalized wavefunction by
$\bar{\psi}(\x) \simeq \bar{\psi}_0(\x)$
where $\bar{\psi}_0$ is the normalized ground-state wavefunction of the potential $V_0(\x)$ (in the absence of disorder).
Plugging this into Eq.~\eqref{V1lambdapsibar}, we obtain
\be
V_{1}\left(\x\right)\simeq\lambda\bar{\psi}_{0}\left(\x\right)^{2} \, .
\ee
The limit $E \to E_0$ corresponds to the limit $\lambda \to 0$. In this limit, $V_1$ is a small perturbation on top of $V_0$. The energy can thus be calculated from $V_1(\x)$ using first-order perturbation theory, which yields
\be
E-E_{0}\simeq\int V_{1}\left(\x\right)\bar{\psi}_{0}\left(\x\right)^{2}d\x \simeq \lambda \int \bar{\psi}_{0}\left(\x\right)^{4}d\x \, .
\ee
On the other hand, the action is given by
\be
s=\frac{1}{2}\int V_{1}\left(\x\right)^{2}d\x \simeq \frac{\lambda^2}{2} \int \bar{\psi}_{0}\left(\x\right)^{4}d\x \, .
\ee
Putting the last two equations together, we can express $s$ as a function of $E$:
\be
\label{sSolTypicalFluc}
s\left(E\right)\simeq\frac{\left(E-E_{0}\right)^{2}}{2\int\bar{\psi}_{0}\left(\x\right)^{4}d\x} \, .
\ee
Therefore, $s(E)$ is in general parabolic around its minimum $E=E_0$, with a coefficient that depends in a simple way on the unperturbed ground-state wave function.
Recalling Eq.~\eqref{LDP}, we find that this parabolic behavior of $s(E)$ corresponds to a Gaussian distribution of typical fluctuations, with mean $\left\langle E\right\rangle \simeq E_{0}$ and variance $\text{Var}(E)\simeq \epsilon\int\bar{\psi}_{0}\left(\x\right)^{4}d\x$.

\subsection{Illustration: Harmonic trap $V_0(\x)$}

Let us illustrate the results above for the harmonic oscillator $V_0(\x) =x^2 \! / 2$, in dimensions $d=1,2,3$.

\textbf{Numerical results:}
We compute $s(E)$ by numerically solving Eq.~\eqref{psiEq} at all $E$ using the shooting method. 
In $d=1$ we use the mirror symmetry $\psi(x) = \psi(-x)$ and solve Eq.~\eqref{psiEq} for $x \ge 0$, with the boundary condition $\psi'(0) = 0$.
In $d>1$ we use the radial symmetry $\psi(\x) = \psi(x)$ so in Eq.~\eqref{psiEq} we replace the Laplacian by 
$\nabla^{2}\psi\left(\x\right)=\frac{1}{x^{d-1}}\frac{d}{dx}\left(x^{d-1}\frac{d\psi}{dx}\right)$.
This yields an effective one-dimensional problem in terms of the radial coordinate $x \ge 0$ which we solve again together with the boundary condition $\psi'(0) = 0$ (which follows from the radial symmetry).
At all $d$, the shooting parameter is $\psi(0)$.

\textbf{Analytic results in limiting cases:}
The leading-order behavior \eqref{sEMinusInf} at $E \to -\infty$ is universal: It depends only on $d$ but not on $V_0(\x)$, under the mild conditions given above (these conditions indeed hold for the harmonic potential $V_0(\x) = x^2 \! /2$).
At $E \to \infty$, Eq.~\eqref{sSolEInf} gives
\bea
\label{sHOEInf1d}
\!\! s\!&\simeq&\!\frac{1}{2}\int_{-\sqrt{2E}}^{\sqrt{2E}}\left(\! E-\frac{x^{2}}{2}\right)^{2}dx=\frac{8\sqrt{2}}{15}E^{5/2},\\[1mm]
\label{sHOEInf2d}
\!\! s\!&\simeq&\!\frac{1}{2}\int_{0}^{\sqrt{2E}} \! \left(\! E-\frac{x^{2}}{2}\right)^{2}2\pi xdx=\frac{\pi}{3}E^{3},\\[1mm]
\label{sHOEInf3d}
\!\! s\!&\simeq&\!\frac{1}{2}\int_{0}^{\sqrt{2E}} \! \left(\! E-\frac{x^{2}}{2}\right)^{2}4\pi x^{2}dx=\frac{32\sqrt{2}\,\pi}{105}E^{7/2}
\eea
in $d=1,2,3$ respectively.

\begin{figure}[h!]
\includegraphics[width=0.48\textwidth,clip=]{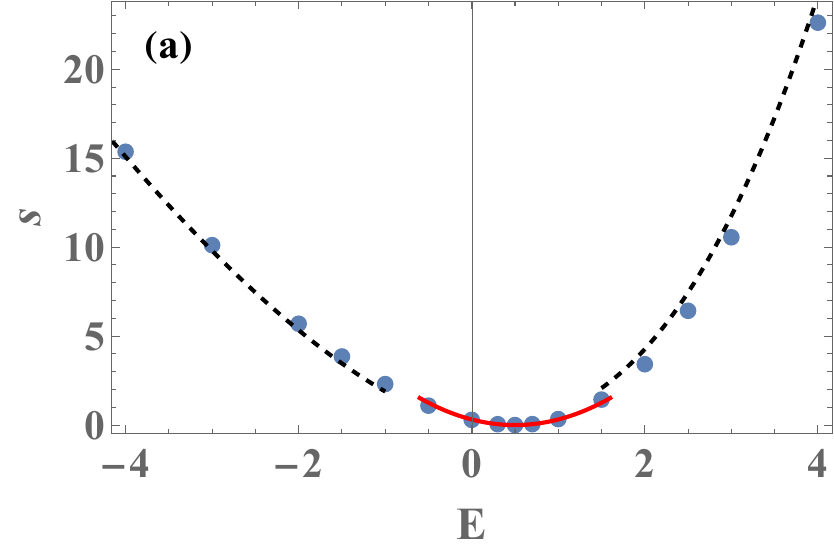}

\vspace{2mm}

\includegraphics[width=0.48\textwidth,clip=]{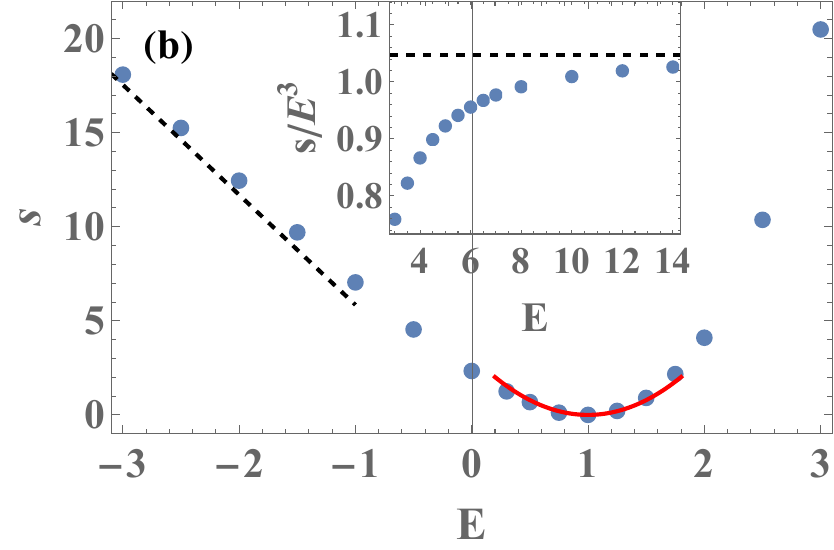}

\vspace{2mm}

\includegraphics[width=0.48\textwidth,clip=]{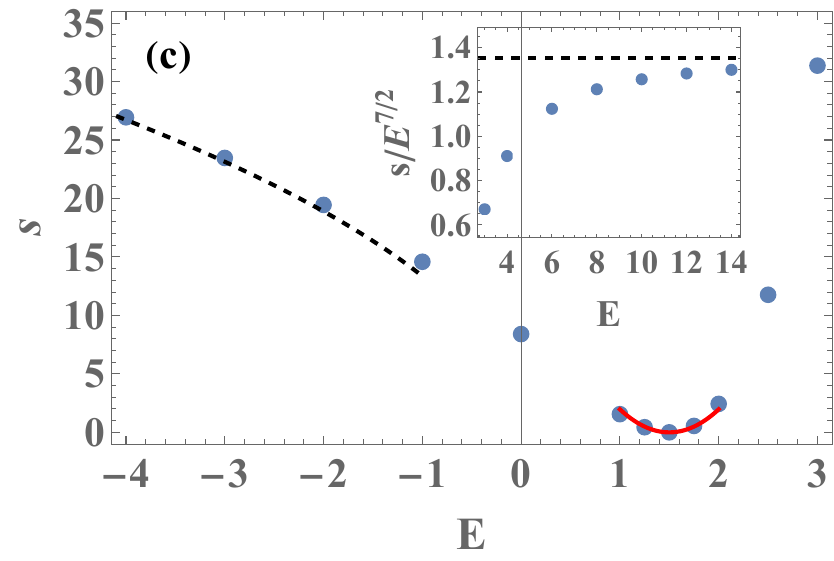}
\caption{The large deviation function $s(E)$ for harmonic trap $V_0(\x) = \x^2\!/2$ in dimensions $d=1,2,3$ [panels (a), (b) and (c) respectively]. Markers are obtained by evaluating the action \eqref{sOfPsi} on numerical solutions to Eq.~\eqref{psiEq}. Solid and dashed lines correspond to the asymptotic behaviors at $E \simeq E_0 = d/2$ and $E \to \pm \infty$, see Eqs.~\eqref{sEMinusInf}, \eqref{sHOEInf1d}-\eqref{sHOEInf3d} and \eqref{sHOTypical}. The insets in (b) and (c) demonstrate the asyptotic convergence in the tail $E \to \infty$.}
\label{figsOfE}
\end{figure}

At $E \simeq E_0$, we use that the normalized ground-state wave function of the harmonic oscillator in the absence of disorder is
\be
\bar{\psi}_{0}\left(\x\right)=\pi^{-d/4}e^{-x^{2} \! /2}
\ee
with corresponding energy $E_0 = d/2$.
Plugging this into Eq.~\eqref{sSolTypicalFluc}, we obtain
\be
\label{sHOTypical}
s\left(E\right)\simeq\frac{2\left(E-E_{0}\right)^{2}}{\pi^{-d}\left(\int_{-\infty}^{\infty}e^{-2x^{2}}dx\right)^{d}}=\frac{\left(2\pi\right)^{d/2}}{2}\left(E-\frac{d}{2}\right)^{2}.
\ee

The asymptotic expressions for $s(E)$ are compared to numerical computations of $s(E)$ in Fig.~\ref{figsOfE}, showing excellent agreement. In $d=2,3$ the convergence in the limit $E \to \infty$ is a little slow, and can be observed in the insets.


\section{Finite, homogenous, periodic system}
\label{sec:PBC}

Let us consider now a finite, one-dimensional system with periodic boundary conditions. We choose units of length such that $-1<x<1$. For simplicity, let us consider a homogeneous background, $V_0(x)=0$.
In the absence of disorder, the ground-state energy is $E_0$ = 0, with a constant associated wave function.
In order to obtain the full distribution of $E$, we must solve Eq.~\eqref{psiEq}, which for our case takes the form
\be
\label{psiEqPBC}
-\frac{1}{2}\psi''\left(x\right)+\text{sgn}\left(E\right)\psi\left(x\right)^{3}=E\psi\left(x\right),
\ee
on the interval $-1<x<1$ with periodic boundary conditions $\psi(x+2) = \psi(x)$.

\begin{figure*}
\includegraphics[width=0.47\textwidth,clip=]{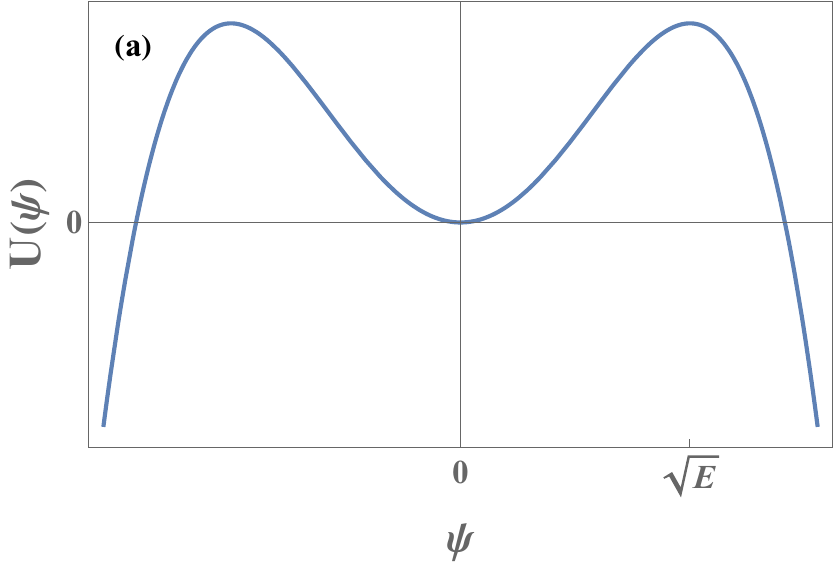}
\hspace{2mm}
\includegraphics[width=0.47\textwidth,clip=]{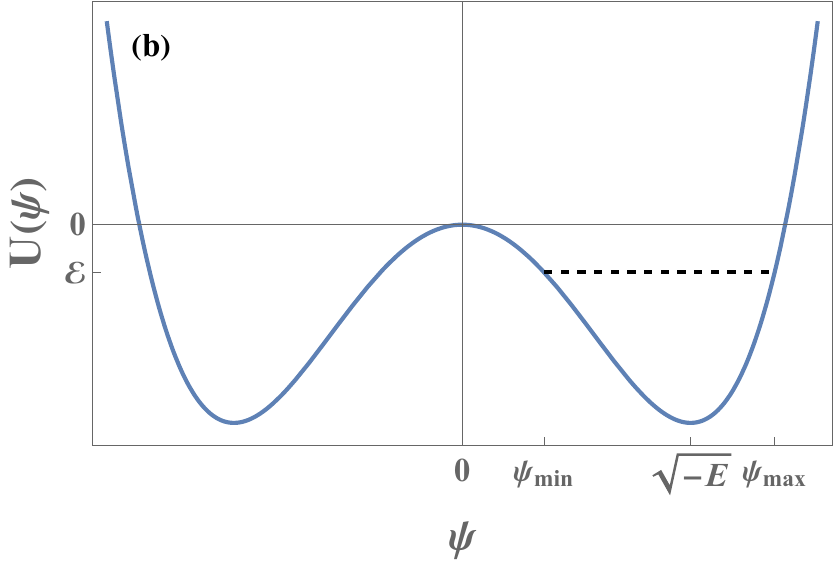}
\caption{The effective potential $U(\psi)$ for $E>0$ (a) and $E<0$ (b) that describes the solutions for the periodic system, see Eq.~\eqref{psiEqPBC2}. The Newtonian dynamics in the potential $U(\psi)$ have nonzero fixed points at $\psi = \pm \sqrt{|E|}$, describing homogeneous solutions. At $E<0$ the dynamics admit inhomogeneous solutions which oscillate between two positive values $\psi_{\min}$ and $\psi_{\min}$, whose mechanical energy $\mathcal{E}$ is denoted by the dashed line in (b).}
\label{figUofPsi}
\end{figure*}

Let us rewrite Eq.~\eqref{psiEqPBC} in the form
\be
\label{psiEqPBC2}
\psi''\left(x\right)=-\frac{dU}{d\psi},\quad U\left(\psi\right)=E\psi^{2}-\frac{1}{2}\text{sgn}\left(E\right)\psi^{4},
\ee
from which a mechanical analogy is immediately apparent. Indeed, Eq.~\eqref{psiEqPBC2} may be interpreted as Newton's second law of motion, where $\psi$ and $x$ play the roles of position and time, respectively, for a particle of unit mass in the potential $U\left(\psi\right)$.
In what follows, it will be useful to recall that the ground-state wave function is real and does not change sign, and thus we will (without loss of generality) choose it to be positive $\psi(x)>0$.

\smallskip
\textbf{Homogeneous solution:}
At all $E$, there exists a homogeneous solution to Eq.~\eqref{psiEqPBC}
\be
\psi_{\text{hom}}\left(x\right)=\sqrt{\left|E\right|} \, .
\ee
Within the mechanical analogy, this solution corresponds to the nonzero extremum of $U(\psi)$, which is a local maximum (minimum) of $U(\psi)$ for $E>0$ ($E<0$), see Fig.~\ref{figUofPsi}.
Recalling Eq.~\eqref{V1ofPsi}, the homogeneous solution simply corresponds to the (quantum) potential
$V(x) = V_1(x) = E$.
The action \eqref{sOfPsi} evaluated on the homogeneous solution is
\be
s_{\text{hom}}=\frac{1}{2}\int_{-1}^{1}\psi_{\text{hom}}\left(x\right)^{4}dx=E^{2} \, .
\ee
The homogeneous solution is indeed the minimizer of the action functional under the constraints for a broad range of energies $E$. However, as we will now show, for sufficiently low energies there exists another solution with lower action.

\smallskip
\textbf{Inhomogeneous solution:}
Inhomogeneous solutions break the translational symmetry of the problem and describe nontrivial motion in the potential $U(\psi)$. Due to the periodic boundary conditions, $\psi(x+2) = \psi(x)$, this motion must take the form of oscillations whose period is of the form $2/k$ where $k$ is a positive integer.
In fact, it turns out that if multiple inhomogeneous solutions exist, the optimal one (i.e., the minimizer of the action constrained on $E$) is the one with $k=1$, i.e., a period of exactly 2.
Since we assume $\psi(x)>0$, these oscillations may not involve $\psi$ crossing the origin. For $E>0$, no such oscillating solutions exist (see Fig.~\ref{figUofPsi}(a)), so for $E>0$ the correct solution is the homogeneous one.

For $E<0$, however, there exist oscillating solutions for which $\psi(x)>0$. Indeed, the conservation of ``mechanical energy'' corresponding to Eq.~\eqref{psiEqPBC2} is
\be
\label{energyConservation}
\frac{1}{2}\psi'\left(x\right)^{2}+U\left(\psi\right)=\frac{1}{2}\psi'\left(x\right)^{2}+E\psi^{2}+\frac{1}{2}\psi^{4}=\mathcal{E}=\text{const},
\ee
 and one finds that for $\mathcal{E} \! < \! 0$, there are oscillating solutions with $\psi(x) \! > \! 0$, see Fig.~\ref{figUofPsi}(b).
The oscillations of $\psi$ as a function of $x$ are between the values $\psi_{\min} \! = \! \sqrt{-E-\sqrt{E^{2}+2\mathcal{E}}}$ and $\psi_{\max} \! = \! \sqrt{-E+\sqrt{E^{2}+2\mathcal{E}}}$, which are the positive solutions to the equation $U\left(\psi\right) = \mathcal{E}$.
Rearranging Eq.~\eqref{energyConservation}, we obtain
\be
\frac{d\psi}{\sqrt{2\mathcal{E}-2E\psi^{2}-\psi^{4}}}=\pm dx \, .
\ee
Assuming now that the period is exactly $2$, we integrate the last equation over half an oscillation. This yields
\be
\label{mathcalEeq}
\int_{\psi_{\min}}^{\psi_{\max}}\frac{d\psi}{\sqrt{2\mathcal{E}-\psi^{4}-2E\psi^{2}}}=1 \, .
\ee
Eq.~\eqref{mathcalEeq} determines the correct value of $\mathcal{E}$ as a function of $E$.

However, it is technically difficult to solve Eq.~\eqref{mathcalEeq} for $\mathcal{E}$, so we will take a different route.
Changing the integration variable, $\psi=\sqrt{-E}\,u$, we rewrite Eq.~\eqref{mathcalEeq} in the form
\be
\label{EofBetaPreSol}
\int_{u_{\min}}^{u_{\max}}\frac{1}{\sqrt{2\beta-u^{4}+2u^{2}}}du=\sqrt{-E},
\ee
where $\beta=\mathcal{E}/E^{2}$ and
\be
u_{\min,\max}=\frac{\psi_{\min,\max}}{\sqrt{-E}}=\sqrt{1\mp\sqrt{1+2\beta}} \, .
\ee
We now solve the integral in Eq.~\eqref{EofBetaPreSol} by rewriting it in the form
\bea
\label{EofBeta}
\sqrt{-E} &=& \int_{u_{\min}}^{u_{\max}}\frac{1}{\sqrt{\left(u^{2}-u_{\min}^{2}\right)\left(u_{\max}^{2}-u^{2}\right)}}du \nn\\
&=&\frac{1}{\sqrt{\sqrt{2\beta+1}+1}}\mathbb{K}\left(\frac{2}{1+\frac{1}{\sqrt{2\beta+1}}}\right) \, ,
\eea
where $\mathbb{K}(\dots)$ is the the complete elliptic integral of the first kind \cite{EllipticKWolfram}.
Next, we calculate the action $s$ by expressing it as a function of $\beta$.
It is sufficient to evaluate the integral \eqref{sOfPsi} over half an oscillation, and then multiply the result by $2$ [which cancels out with the factor of $1/2$ in \eqref{sOfPsi}]. Then, changing the integration variable from $x$ to $\psi$, one obtains
\be
s=\int_{\psi_{\min}}^{\psi_{\max}}\frac{\psi^{4}d\psi}{\left|\psi'\left(x\right)\right|}=\int_{\psi_{\min}}^{\psi_{\max}}\frac{\psi^{4}d\psi}{\sqrt{2\mathcal{E}-\psi^{4}-2E\psi^{2}}} \, .
\ee
Changing variables again $\psi=\sqrt{-E}\,u$, we obtain
\begin{widetext}
\bea
\label{sofBeta}
s\left(\beta\right)&=&\left(-E\left(\beta\right)\right)^{3/2}\int_{u_{\min}}^{u_{\max}}\frac{u^{4}}{\sqrt{\left(u^{2}-u_{\min}^{2}\right)\left(u_{\max}^{2}-u^{2}\right)}}du\nn\\[1mm]
&=&\left(-E\left(\beta\right)\right)^{3/2}\frac{\sqrt{\sqrt{2\beta+1}+1}}{3}\left[4\mathbb{E}\left(\frac{2}{1+\frac{1}{\sqrt{2\beta+1}}}\right)+\left(\sqrt{2\beta+1}-1\right)\mathbb{K}\left(\frac{2}{1+\frac{1}{\sqrt{2\beta+1}}}\right)\right]
\eea
\end{widetext}
where $\mathbb{E}(\dots)$ is the the complete elliptic integral of the second kind \cite{EllipticEWolfram}.

\begin{figure}
\includegraphics[width=0.48\textwidth,clip=]{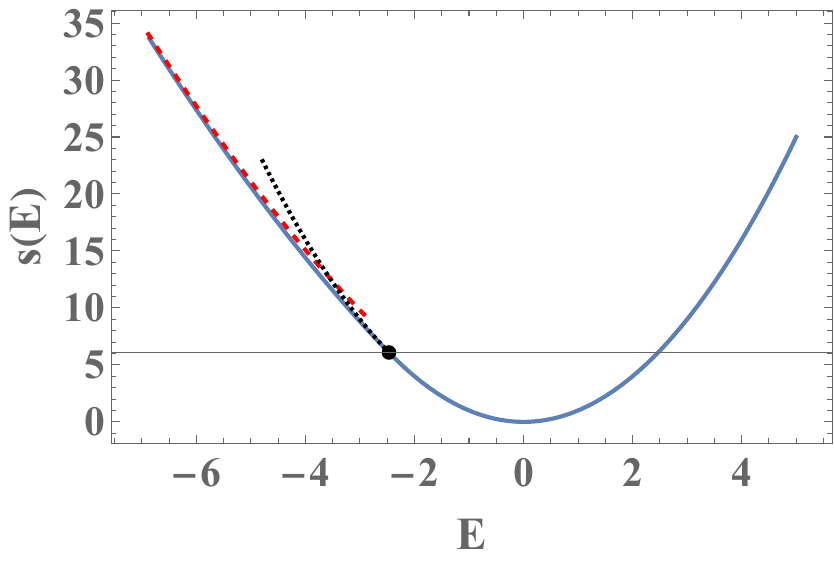}
\caption{Solid line: The large deviation function $s(E)$ for a finite system $-1 < x < 1$ with periodic boundary conditions and no external potential, $V_0(x)=0$. The fat dot marks the critical point $E_c=-\pi^2/4$ at which $s(E)$ is nonanalytic, and which separates between the homogeneous phase $E>E_c$ and the inhomogeneous phase $E<E_c$. The dashed line is the approximate expression \eqref{sMinusInfPBC} at $-E \gg 1$, which coincides with that of an infinite system. The dotted line is the action $s$ evaluated on the homogenous (non-optimal) solution at $E<E_c$, which is higher than that of the (optimal) inhomogeneous solution.}
\label{figsOfEPBC}
\end{figure}

Eqs.~\eqref{EofBeta} and \eqref{sofBeta} determine $s$ as a function of $E$ in a parametric form for the inhomogeneous solution, where the parameter $\beta$ ranges between the values $-1/2 < \beta < 0$.
Interestingly, the inhomogeneous solution only exists for $E < E_c$ where the critical value is $E_c = -\pi^2 / 4$.
One finds that, when the inhomogeneous solution exists, its action (for a given $E$) is indeed smaller than that of the homogeneous solution.
In Fig.~\ref{figsOfEPBC} the large-deviation function $s(E)$ is plotted. It is given by the homogeneous solution for $E>E_c$, and by the inhomogeneous one for $E<E_c$.

Let us now discuss the asymptotic behaviors of $s(E)$.
The limit $E \to E_c$ (from below) corresponds (in the inhomogeneous solution) to the limit $\beta \to -1/2$, or $\mathcal{E} \to -E^2/2=U(\sqrt{-E})$,
i.e., the mechanical energy is only slightly larger than the minimum of the effective potential.
The asymptotic expansions of Eqs.~\eqref{EofBeta} and \eqref{sofBeta} at $\beta \to -1/2$ yield
\bea
\! E&\simeq&-\frac{\pi^{2}}{4}-\frac{3\pi^{2}}{16}\left(\! \beta+\frac{1}{2}\right)-\frac{123\pi^{2}}{512}\left(\! \beta+\frac{1}{2}\right)^{2}  ,\\
\! s&\simeq&\frac{\pi^{4}}{16}+\frac{3\pi^{4}}{32}\left(\beta+\frac{1}{2}\right)+\frac{135\pi^{4}}{1024}\left(\beta+\frac{1}{2}\right)^{2}
\eea
from which we obtain, for $E$ slightly smaller than $E_c$,
\be
s\left(E\right)\simeq\frac{\pi^{4}}{16}-\frac{\pi^{2}}{2}\left(E-E_{c}\right)+\frac{1}{3}\left(E-E_{c}\right)^{2},\;\;E_{c}-E\ll1.
\ee
On the other hand, at $E>E_c$, one has
\be
s\left(E\right)=E^{2}=\frac{\pi^{4}}{16}-\frac{\pi^{2}}{2}\left(E-E_{c}\right)+\left(E-E_{c}\right)^{2},\;\;E>E_{c}.
\ee
We thus find that the second derivative $d^2 s / dE^2$ is discontinuous at $E=E_c$. 
In analogy with thermodynamics in equilibrium, we interpret this nonanalytic behavior of $s(E)$ as a second-order DPT \cite{Baek15, Baek17, Baek18,  Shpielberg2016, SKM2018, ZM16, SMS2018, Agranov2020, Smith22KPZ, Smith22Chaos, ACJ22, SGG23, ACJ23, ME23, SSM23}, where here $s(E)$ plays the role of a nonequilibrium free energy.
In the limit $-E \gg 1$, using the asymptotic behavior \cite{NISTEK}
\be
\mathbb{K}(x) \simeq \frac{1}{2}\ln\left(\frac{1}{1-x}\right)+2\ln2 \, 
\ee
and $\mathbb{E}(1)=1$, Eq.~\eqref{sofBeta} simplifies to
\be
\label{sMinusInfPBC}
s\simeq\frac{4\sqrt{2}}{3}\left(-E\right)^{3/2}
\ee
coinciding with the result for an infinite system \cite{GH24}, i.e., with the first line in \eqref{sEMinusInf}, see also Appendix \ref{app:EMinusInf}.

%
%
%
%

\section{Discussion}
\label{sec:discussion}

To summarize, we have calculated the full distribution (including large deviations) $P(E)$ of the ground-state energy $E$ of a single particle in a potential that is the sum of a ``background'' $V_0(\x)$ and a white-noise disorder  $V_1(x) = \sqrt{\epsilon} \, v_1(\x)$, in the weak-disorder limit $\epsilon \to 0$.
We calculated the large-deviation function $s(E)$ that describes the distribution in the limits $E \to \pm \infty$ and $E \simeq E_0$, where $E_0$ is the ground-state energy in the absence of disorder, for arbitrary dimension $d$ and $V_0(\x)$.
We found that the tail $E \to -\infty$ is universal: It depends only on $d$ but not on $V_0(\x)$, provided that $V_0(\x)$ has a single global minimum $V_0(0)=0$. Moreover, at $E \to -\infty$, $P(E)$ coincides with the density of low-lying states.
We illustrated our results for the harmonic background $V_0(\x) \propto x^2$ in dimensions $d=1,2,3$.
Furthermore, we studied a periodic, one-dimensional homogeneous ($V_0(x)=0$) system and found that a DPT occurs at a critical value $E=E_c$, separating between a homogeneous phase $E>E_c$ and inhomogeneous phase $E<E_c$.

It would be interesting to study the subleading order in small $\epsilon$ that gives the pre-exponential factor which is the subleading correction to the scaling behavior \eqref{LDP}. One way in to achieve this is by considering small fluctuations around the optimal realization of $V(\x)$. Such results have been obtained for the DOS for disordered potentials \cite{GH24}, and also in several other contexts in which the OFM has been applied \cite{KLP18, SGG21, NT22, SGG23}.
For arbitrary (not necessarily small) $\epsilon$, our theory is still expected to describe the distribution correctly  (in the leading order) if one goes sufficiently far into the tails $E \to \pm \infty$. It would be interesting to study the typical-fluctuations regime for arbitrary $\epsilon$.

We focused on white-noise disorder, and it would be interesting to extend our results to other types of disorder, such as spatially-correlated (``colored'') noise.
It would be interesting to study other spectral properties of disordered potentials beyond the ground-state energy (e.g., the statistics of spectral gaps, ionization energies etc).
Finally it would be useful, but challenging, to extend the investigations to the setting of a many body system in an external potential.

\bigskip

\section*{ACKNOWLEDGMENTS}

I thank Baruch Meerson, Pierre Le Doussal, Satya Majumdar and Alex Kamenev for illuminating discussions and for useful comments on the manuscript.
I acknowledge support from the Israel Science Foundation (ISF) through Grant No. 2651/23, and from the Golda Meir Fellowship.

\bigskip

\appendix

\section{Analytic solution in the limit $E \to -\infty$ for $d=1$}
\label{app:EMinusInf}
\renewcommand{\theequation}{A\arabic{equation}}
\setcounter{equation}{0}

Our starting point is Eq.~\eqref{psiEqNoV0}, which in $d=1$ reads
\be
-\frac{1}{2}\psi''\left(x\right)-\psi\left(x\right)^{3}=E\psi\left(x\right)\,.
\ee
The equation is to be solved under the boundary conditions $\psi\left(x \to \pm\infty\right) = 0$.
Following the mechanical analogy described in Section \ref{sec:PBC}, we analyze the motion of a classical particle in the effective potential
$U\left(\psi\right)=E\psi^{2}+\frac{1}{2}\psi^{4}$
[which is the expression in \eqref{psiEqPBC2} for the case $E<E_0$],
see Fig.~\ref{figUofPsi}.
The trajectory must begin and end at the origin, and therefore the ``mechanical energy''  must vanish, $\mathcal{E}=0$.
Thus, the ``energy conservation'' equation \eqref{energyConservation} takes the form
\be
\label{energyConservation0}
\frac{1}{2}\psi'\left(x\right)^{2}+U\left(\psi\right)=\frac{1}{2}\psi'\left(x\right)^{2}+E\psi^{2}+\frac{1}{2}\psi^{4}=0 \, ,
\ee
leading to
\be
\frac{d\psi}{\sqrt{-2E\psi^{2}-\psi^{4}}}=\pm dx \, .
\ee
Integrating the last equation, we obtain
\be
\frac{1}{\sqrt{-2E}}\coth^{-1}\left(\sqrt{\frac{2E}{2E+\psi^{2}}}\,\right) = \pm (x-x_0).
\ee
Solving for $\psi$, we obtain
\be
\label{psiSolEMnusInf}
\psi\left(x\right)=\frac{\sqrt{-2E}}{\cosh\left(\sqrt{-2E}\,\left(x-x_{0}\right)\right)}\,.
\ee
Eq.~\eqref{psiSolEMnusInf} describes a family of solutions, which are all related to each other through translations.
Evaluating the action \eqref{sOfPsi} on this solution, we find
\bea
\!\! s\!&=&\!\frac{1}{2}\int_{-\infty}^{\infty}\frac{4E^{2}}{\cosh^{4}\left(\sqrt{-2E}\,\left(x-x_{0}\right)\right)}dx\nn\\
\!\! &=&\!\sqrt{2}\left(-E\right)^{3/2} \! \int_{-\infty}^{\infty}\frac{dy}{\cosh^{4}y}=\frac{4\sqrt{2}}{3}\left(-E\right)^{3/2}  ,
\eea
coinciding with the first line of Eq.~\eqref{sEMinusInf}.
Finally, Eq.~\eqref{psiSolEMnusInf} describes a solution that is strongly localized around the point $x=x_0$, over a spatial region of order $1/\sqrt{-E} \ll 1$. This justifies the approximation $V(x) \simeq V(0) = 0$, where the choice $x_0=0$ follows from the assumption that the minimum of the potential is at $x=0$. For a finite system, the localization allows one to ignore the boundary conditions and approximate the system as infinite.


\bigskip\bigskip

\begin{thebibliography}{99}

\bibitem{Halperin65} B. I. Halperin, \textit{Green's Functions for a Particle in a One-Dimensional Random Potential}, {\href{https://journals.aps.org/pr/abstract/10.1103/PhysRev.139.A104}{Phys. Rev. \textbf{139}, A104 (1965)}}.

\bibitem{HalperinLax66} B. I. Halperin and M. Lax, \textit{Impurity-Band Tails in the High-Density Limit. I. Minimum Counting Methods}, {\href{https://journals.aps.org/pr/abstract/10.1103/PhysRev.148.722}{Phys. Rev. \textbf{148}, 722 (1966)}}.

\bibitem{ZL66} J. Zittartz and J. S. Langer, \textit{Theory of Bound States in a Random Potential}, {\href{https://journals.aps.org/pr/abstract/10.1103/PhysRev.148.741}{Phys. Rev. \textbf{148}, 741 (1966)}}.

\bibitem{VV95} O.K. Vorov and A.V. Vagov, \textit{Problem of a quantum particle in a random potential on a line revisited}, {\href{https://www.sciencedirect.com/science/article/pii/037596019592832O}{Phys. Lett. A \textbf{205}, 301 (1995)}}.

\bibitem{RoatiEtAl08} G. Roati, C. D'Errico, L. Fallani, M. Fattori, C. Fort, M. Zaccanti, G. Modugno, M. Modugno, and M. Inguscio, \textit{Anderson localization of a non-interacting Bose–Einstein condensate} {\href{https://www.nature.com/articles/nature07071}{Nature (London) \textbf{453}, 895 (2008)}}.

\bibitem{BW12} M. Bishop and J. Wehr, \textit{Ground state energy of the one-dimensional discrete random Schrödinger operator with Bernoulli potential}, {\href{https://link.springer.com/article/10.1007/s10955-012-0480-3}{J. Stat. Phys. \textbf{147}, 529 (2012)}}.

\bibitem{ArnoldEtAl16} D. N. Arnold, G. David, D. Jerison, S. Mayboroda, and M. Filoche, \textit{Effective Confining Potential of Quantum States in Disordered Media}, {\href{https://journals.aps.org/prl/abstract/10.1103/PhysRevLett.116.056602}{Phys. Rev. Lett. \textbf{116}, 056602 (2016)}}. 

\bibitem{CWZ22} I. Chenn, W. Wang and S. Zhang, \textit{Approximating the ground state eigenvalue via the effective potential}, {\href{https://iopscience.iop.org/article/10.1088/1361-6544/ac692a}{Nonlinearity \textbf{35}, 3004 (2022)}}.

\bibitem{GH24} E. R. Garcia and J. Hofmann, \textit{Fluctuation corrections to Lifshitz tails in disordered systems}, {\href{https://journals.aps.org/pre/abstract/10.1103/PhysRevE.109.L032103}{Phys. Rev. E \textbf{109}, L032103 (2024)}}. 

\bibitem{VanMieghem92} P. Van Mieghem, \textit{Theory of band tails in heavily doped semiconductors}, {\href{https://journals.aps.org/rmp/abstract/10.1103/RevModPhys.64.755}{Rev. Mod. Phys. \textbf{64}, 755 (1992)}}. 

\bibitem{Affleck83} I. Affleck, \textit{Two-dimensional disorder in the presence of a uniform magnetic field}, {\href{https://iopscience.iop.org/article/10.1088/0022-3719/16/30/014}{J. Phys. C: Solid State Phys. \textbf{16}, 5839 (1983)}}.

\bibitem{JeanEtAl17} J. Jean, T. S. Mahony, D. Bozyigit, M. Sponseller, J. Holovský, M. G. Bawendi, and V. Bulovi\'{c}, \textit{Radiative Efficiency Limit with Band Tailing Exceeds 30\% for Quantum Dot Solar Cells}, {\href{https://pubs.acs.org/doi/10.1021/acsenergylett.7b00923}{ACS Energy Lett. \textbf{2}, 2616 (2017)}}.

\bibitem{WOA21} J. Wong, S. T. Omelchenko, and H. A. Atwater, \textit{Impact of Semiconductor Band Tails and Band Filling on Photovoltaic Efficiency Limits}, {\href{https://pubs.acs.org/doi/10.1021/acsenergylett.0c02362}{ACS Energy Lett. \textbf{6}, 52 (2021)}}.

\bibitem{FPY96} V. Freilikher, M. Pustilnik, and I. Yurkevich, \textit{Enhanced transmission through a disordered potential barrier}, {\href{https://journals.aps.org/prb/abstract/10.1103/PhysRevB.53.7413}{Phys. Rev. B \textbf{53}, 7413 (1996)}}.

\bibitem{DaCostaEtAlExperiment2000} V. Da Costa, Y. Henry, F. Bardou, M. Romeo, and K. Ounadjela, \textit{Experimental evidence and consequences of rare events in quantum tunneling}, {\href{https://doi.org/10.1007/s100510050035}{Eur. Phys. J. B \textbf{13}, 297 (2000)}}.

\bibitem{Luck04} J. M. Luck, \textit{Non-monotonic disorder-induced enhanced tunnelling}, {\href{https://iopscience.iop.org/article/10.1088/0305-4470/37/1/018}{J. Phys. A: Math. Gen. \textbf{37}, 259 (2004)}}.

\bibitem{KRL08} K. Kim, F. Rotermund, and H. Lim, \textit{Disorder-enhanced transmission of a quantum mechanical particle through a disordered tunneling barrier in one dimension: Exact calculation based on the invariant imbedding method}, {\href{https://journals.aps.org/prb/abstract/10.1103/PhysRevB.77.024203}{Phys. Rev. B \textbf{77}, 024203 (2008)}}.

\bibitem{BillyEtAl08} J. Billy, V. Josse, Z. Zuo, A. Bernard, B. Hambrecht, P. Lugan, D. Clément, L. Sanchez-Palencia, P. Bouyer, and A. Aspect, \textit{Direct observation of Anderson localization of matter waves in a controlled disorder}, {\href{https://www.nature.com/articles/nature07000}{Nature (London) \textbf{453}, 891 (2008)}}.


\bibitem{Yaida16} S. Yaida, \textit{Instanton calculus of Lifshitz tails}, {\href{https://journals.aps.org/prb/abstract/10.1103/PhysRevB.93.075120}{Phys. Rev. B \textbf{93}, 075120 (2016)}}.





\bibitem{Varadhan} S. S. Varadhan, \textit{Large Deviations and Applications}, CBMS-NSF Regional Conference Series in Applied Mathematics, No. 46 (SIAM, Philadelphia, 1984).
\bibitem{O1989} Y. Oono, \textit{Large Deviation and Statistical Physics}, {\href{https://academic.oup.com/ptps/article/doi/10.1143/PTPS.99.165/1880363}{Prog. Theor. Phys. Suppl. \textbf{99}, 165 (1989)}}.
\bibitem{DZ} A. Dembo and O. Zeitouni, \textit{Large Deviations Techniques
and Applications}, 2nd ed. (Springer, New York, 1998).
\bibitem{Hollander} F. den Hollander, \textit{Large Deviations}, Fields Institute Monographs, vol. 14 (AMS, Providence, Rhode Island, 2000).




\bibitem{Derrida2007} B. Derrida, \textit{Non-equilibrium steady states: fluctuations and large deviations of the density and of the current}, {\href{https://iopscience.iop.org/article/10.1088/1742-5468/2007/07/P07023}{J. Stat. Mech. (2007) P07023.}}

\bibitem{T2009} H. Touchette, \textit{The large deviation approach to statistical mechanics}, {\href {https://www.sciencedirect.com/science/article/pii/S0370157309001410?via\%3Dihub}{Phys. Rep. \textbf{478}, 1 (2009).}}


\bibitem{Derrida11} B. Derrida, \textit{Microscopic versus macroscopic approaches to non-equilibrium systems}, {\href{https://iopscience.iop.org/article/10.1088/1742-5468/2011/01/P01030/meta}{J. Stat. Mech. (2011) P01030}}.




\bibitem{bertini2015} L. Bertini, A. De Sole, D. Gabrielli, G. Jona-Lasinio, and C. Landim, \textit{Macroscopic fluctuation theory}, {\href{https://journals.aps.org/rmp/abstract/10.1103/RevModPhys.87.593}{Rev. Mod. Phys. \textbf{87}, 593 (2015)}}.



\bibitem{Baek15}   Y. Baek and Y. Kafri, \textit{Singularities in large deviation functions
}, {\href{https://iopscience.iop.org/article/10.1088/1742-5468/2015/08/P08026}{J. Stat. Mech. (2015) P08026}}.
\bibitem{ZM16} L. Zarfaty, B. Meerson, \textit{Statistics of large currents in the Kipnis-Marchioro-Presutti model in a ring geometry}, {\href{https://iopscience.iop.org/article/10.1088/1742-5468/2016/03/033304}{J. Stat. Mech. (2016) 033304}}.
\bibitem{Shpielberg2016} O. Shpielberg and E. Akkermans, \textit{Le Chatelier Principle for Out-of-Equilibrium and Boundary-Driven Systems: Application to Dynamical Phase Transitions}, {\href{https://journals.aps.org/prl/abstract/10.1103/PhysRevLett.116.240603}{Phys. Rev. Lett. \textbf{116}, 240603 (2016)}}. 
\bibitem{Baek17}   Y. Baek, Y. Kafri, and V. Lecomte, \textit{Dynamical Symmetry Breaking and Phase Transitions in Driven Diffusive Systems}, {\href{https://journals.aps.org/prl/abstract/10.1103/PhysRevLett.118.030604}{Phys. Rev. Lett. \textbf{118}, 030604 (2017)}}.
\bibitem{Baek18}   Y. Baek, Y. Kafri, and V. Lecomte, \textit{Dynamical phase transitions in the current distribution of driven diffusive channels}, {\href{https://iopscience.iop.org/article/10.1088/1751-8121/aaa8f9/meta} {J. Phys. A: Math. Theor. \textbf{51}, 10500 (2018)}}.



%
\bibitem{MS2017} S. N. Majumdar and G. Schehr, \textit{Large deviations}, ICTS Newsletter 2017 (Volume 3, Issue 2); arXiv:1711.07571.
\bibitem{Touchette2018} H. Touchette, \textit{Introduction to dynamical large deviations of Markov processes}, {\href{https://doi.org/10.1016/j.physa.2017.10.046}{Physica A \textbf{504}, 5 (2018)}}.

\bibitem{SMS2018} N. R. Smith, B. Meerson and P. V. Sasorov, \textit{Finite-size effects in the short-time height distribution of the Kardar-Parisi-Zhang equation}, {\href{https://iopscience.iop.org/article/10.1088/1742-5468/aaa783}{J. Stat. Mech. (2018) 023202}}. 


\bibitem{SKM2018} N. R. Smith, A. Kamenev and B. Meerson, \textit{Landau theory of the short-time dynamical phase transitions of the Kardar-Parisi-Zhang interface}, {\href{https://journals.aps.org/pre/abstract/10.1103/PhysRevE.97.042130}{Phys. Rev. E \textbf{97}, 042130 (2018).}} 




\bibitem{KLP18} A. Krajenbrink, P. Le Doussal and S. Prolhac, \textit{Systematic time expansion for the Kardar-Parisi-Zhang equation, linear statistics of the GUE at the edge and trapped fermions},  {\href{https://www.sciencedirect.com/science/article/pii/S0550321318302669}{Nuclear Physics B \textbf{936}, 239 (2018)}}.


\bibitem{Agranov2020} T. Agranov, P. Zilber, N. R. Smith, T. Admon, Y. Roichman, B. Meerson, \textit{The Airy distribution: experiment, large deviations and additional statistics}, {\href{https://journals.aps.org/prresearch/abstract/10.1103/PhysRevResearch.2.013174}{Phys. Rev. Res. \textbf{2}, 013174 (2020)}}.


\bibitem{SGG21} T. Schorlepp, T. Grafke, and R. Grauer, \textit{Gel'fand-Yaglom type equations for calculating fluctuations around instantons in stochastic systems}, {\href{https://iopscience.iop.org/article/10.1088/1751-8121/abfb26}{J. Phys. A: Math. Theor. \textbf{54} 235003 (2021)}}.


\bibitem{NT22} D. Nickelsen, H. Touchette, \textit{Noise correction of large deviations with anomalous scaling}, {\href{https://journals.aps.org/pre/abstract/10.1103/PhysRevE.105.064102}{Phys. Rev. E \textbf{105}, 064102 (2022)}}. 


\bibitem{Smith22KPZ} N. R. Smith, \textit{Exact short-time height distribution and dynamical phase transition in the relaxation of a Kardar-Parisi-Zhang interface with random initial condition}, {\href{https://journals.aps.org/pre/abstract/10.1103/PhysRevE.106.044111}{Phys. Rev. E \textbf{106}, 044111 (2022)}}. 

\bibitem{Smith22Chaos} N. R. Smith, \textit{Large deviations in chaotic systems: Exact results and dynamical phase transition}, {\href{https://journals.aps.org/pre/abstract/10.1103/PhysRevE.106.L042202}{Phys. Rev. E \textbf{106}, L042202 (2022)}}. 


\bibitem{ACJ22} T. Agranov, M. E. Cates, R. L. Jack, \textit{Entropy production and its large deviations in an active lattice gas}, {\href{https://iopscience.iop.org/article/10.1088/1742-5468/aca0eb}{J. Stat. Mech. (2022) 123201}}. 

\bibitem{SGG23} T. Schorlepp,T. Grafke and R. Grauer, \textit{Symmetries and Zero Modes in Sample Path Large Deviations}, {\href{https://link.springer.com/article/10.1007/s10955-022-03051-w}{J. Stat. Phys. \textbf{190}, 50 (2023)}}. 

\bibitem{ACJ23} T. Agranov, M. E. Cates, R. L. Jack, \textit{Tricritical behavior in dynamical phase transitions}, {\href{https://doi.org/10.1103/PhysRevLett.131.017102}{Phys. Rev. Lett. \textbf{131}, 017102 (2023)}}.

\bibitem{ME23} J. Meibohm and M. Esposito, \textit{Landau theory for finite-time dynamical phase transitions}, {\href{https://iopscience.iop.org/article/10.1088/1367-2630/acbc41}{New J. Phys. \textbf{25}, 023034 (2023)}}.

\bibitem{SSM23} T. Schorlepp, P. Sasorov, B. Meerson, \textit{Short-time large deviations of the spatially averaged height of a KPZ interface on a ring}, {\href{https://iopscience.iop.org/article/10.1088/1742-5468/ad0a94}{J. Stat. Mech. (2023) 123202}}. 














\bibitem{Lifshitz} I. M. Lifshitz, Zh. Eksp. Teor. Fiz. \textbf{53}, 743 (1967) [Sov. Phys.
JETP \textbf{26}, 462 (1968)].
\bibitem{turb1} G. Falkovich, I. Kolokolov, V. Lebedev, and A. Migdal, \textit{Instantons and intermittency}, {\href{https://journals.aps.org/pre/abstract/10.1103/PhysRevE.54.4896}{Phys. Rev. E \textbf{54}, 4896 (1996)}}.
\bibitem{EK}  V. Elgart and A. Kamenev, \textit{Rare event statistics in reaction-diffusion systems}, {\href{https://journals.aps.org/pre/abstract/10.1103/PhysRevE.70.041106}{Phys. Rev. E \textbf{70}, 041106 (2004)}}.
\bibitem{MS2011} B. Meerson and P.V. Sasorov, \textit{Extinction rates of established spatial populations}, {\href{https://journals.aps.org/pre/abstract/10.1103/PhysRevE.83.011129}{Phys. Rev. E \textbf{83}, 011129 (2011)}}; \textit{Negative velocity fluctuations of pulled reaction fronts}, {\href{https://journals.aps.org/pre/abstract/10.1103/PhysRevE.84.030101}{\textbf{84}, 030101(R) (2011)}}.
\bibitem{MKV} B. Meerson, E. Katzav, and A. Vilenkin, \textit{Large Deviations of Surface Height in the Kardar-Parisi-Zhang Equation}, {\href{https://journals.aps.org/prl/abstract/10.1103/PhysRevLett.116.070601}{Phys. Rev. Lett. \textbf{116}, 070601 (2016)}}.


\bibitem{LV01} A. V. Lopatin and V. M. Vinokur, \textit{Instanton Approach to the Langevin Motion of a Particle in a Random Potential}, {\href{https://journals.aps.org/prl/abstract/10.1103/PhysRevLett.86.1817}{Phys. Rev. Lett. 86, 1817 (2001)}}.

\bibitem{Meerson22} B. Meerson, \textit{Negative autocorrelations of disorder strongly suppress thermally activated particle motion in short-correlated quenched Gaussian disorder potentials}, {\href{https://journals.aps.org/pre/abstract/10.1103/PhysRevE.105.034106}{Phys. Rev. E \textbf{105}, 034106 (2022)}}; {\href{https://journals.aps.org/pre/abstract/10.1103/PhysRevE.107.039902}{\textbf{107}, 039902(E) (2023)}}; arXiv:2111.07729

\bibitem{VLM24} A. Valov, N. Levi, B. Meerson, \textit{Thermally activated particle motion in biased correlated Gaussian disorder potentials}, {\href{https://journals.aps.org/pre/abstract/10.1103/PhysRevE.110.024138}{Phys. Rev. E \textbf{110}, 024138 (2024)}}. 

\bibitem{footnote:DPT} Although we do not consider dynamics in this paper, we use the term ``dynamical phase transition'' because this is the standard large-deviation theory term used when describing singularities in large deviaiton functions.


\bibitem{Zwanzig88} R. Zwanzig, \textit{Diffusion in a rough potential}, {\href{https://www.pnas.org/doi/10.1073/pnas.85.7.2029}{Proc. Natl. Acad. Sci. USA 85, 2029 (1988)}}.


\bibitem{SB80} J. Stoer and R. Bulirsch, \textit{Introduction to Numerical Analysis} (Springer-Verlag, New York, 1980).



%
%


















\bibitem{CGT64} R. Y. Chiao, E. Garmire and C. H. Townes, \textit{Self-Trapping of Optical Beams}, {\href{https://journals.aps.org/prl/abstract/10.1103/PhysRevLett.13.479}{Phys. Rev. Lett. \textbf{13} 479 (1964)}}.

\bibitem{Coleman77} S. Coleman, \textit{Fate of the false vacuum: Semiclassical theory}, {\href{https://journals.aps.org/prd/abstract/10.1103/PhysRevD.15.2929}{Phys. Rev. D \textbf{15}, 2929 (1977)}}.

\bibitem{MSV_3d} B. Meerson, P. V. Sasorov and A. Vilenkin, \textit{Nonequilibrium steady state of a weakly-driven Kardar–Parisi–Zhang equation}, {\href{https://iopscience.iop.org/article/10.1088/1742-5468/aabbcc}{J. Stat. Mech. (2018) 053201}}. 

\bibitem{footnote:units} The relation between our units and those of Ref.~\cite{GH24} is as follows: Our $\epsilon$ and their $w$ are related via $\epsilon=w^{2}/4$, and our energy $E$ equals half of theirs. As a result, their Eq.~(4) becomes, in the leading order at $E \to -\infty$, our Eq.~\eqref{sEMinusInf}, where our numerical coefficients are related to theirs through
$5.85 = 11.70 / 2$ and $13.36\simeq37.79 / \sqrt{8}$.

\bibitem{Castin2}Y. Castin, arXiv:cond-mat/0407118,
in {\it Quantum gases in low dimensions}, 
J. Phys. IV France, {\bf 116} 89 (2004).

\bibitem{Castin}Y. Castin, 
arXiv:cond-mat/0612613, in {\it Ultra-cold Fermi Gases}, ed. by
M. Inguscio, W. Ketterle, and C. Salomon, (2006).


\bibitem{EllipticKWolfram} \url{https://mathworld.wolfram.com/CompleteEllipticIntegraloftheFirstKind.html}

\bibitem{EllipticEWolfram} \url{https://mathworld.wolfram.com/CompleteEllipticIntegraloftheSecondKind.html}

\bibitem{NISTEK} \url{https://dlmf.nist.gov/19.12}


\end{thebibliography}
\end{document}